\documentclass{aastex631}

\shorttitle{LAMOST J0240+19}
\shortauthors{Barrett}

\graphicspath{{./}{figures/}}

\begin{document}

\title{VLA Observations of the AE~Aqr-type Cataclysmic Variable LAMOST
  J024048.51+195226.9}

\author[0000-0002-8456-1424]{Paul E. Barrett} \affiliation{Department
  of Physics, The George Washington University, Washington, DC 20052
  USA}

\begin{abstract}

  AE~Aqr was until recently the only known magnetic cataclysmic
  variable (MCV) containing a rapidly spinning (33.08 s) white dwarf
  (WD). Its radio emission is believed to be a superposition of
  synchrotron emitting plasmoids, because it has a positive spectral
  index spanning three orders of magnitude ($\approx 2-2000$ GHz) and
  is unpolarized. Both characteristics are unusual for MCVs.
  Recently, Thorstensen has suggested that the cataclysmic variable
  LAMOST J024048.51+195226.9 (henceforth, J0240+19) is a twin of
  AE~Aqr based on its optical spectra. Optical photometry shows the
  star to be a high-inclination, eclipsing binary with a spin period
  of 24.93 s, making it the fastest spinning WD. This paper presents
  three hours of Very Large Array radio observations of
  J0240+19. These observations show that the persistent radio emission
  from J0240+19 is dissimilar to that of AE~Aqr in that it shows high
  circular polarization and a negative spectral index. The emission is
  most similar to the nova-like CV V603~Aql. We argue that the radio
  emission is caused by a superposition of plasmoids emitting plasma
  radiation or electron cyclotron maser emission from the lower corona
  of the donor star and not from the magnetosphere near the WD,
  because the latter site is expected to be modulated at the orbital
  period of the binary and to show eclipses of which there is no
  evidence. The radio source J0240+19, although weak ($\lesssim 1$
  mJy), is a persistent source in a high-inclination eclipsing binary,
  making it a good laboratory for studying radio emission from CVs.

\end{abstract}


\section{Introduction}

Magnetic cataclysmic variables (MCVs) showing two photometric periods
are called intermediate polars (also known as DQ Herculis stars). The
longer period (of order a few hours) is associated with the orbital
period $P_{orb}$ of the binary and the shorter period (of order tens
of minutes) with the rotation period $P_{rot}$ of the white dwarf
(WD). The ratio $P_{rot} / P_{orb} \approx 0.1$.  The exception to the
spin-orbit rule is AE Aquarii (AE~Aqr), which until recently,
contained the fastest known rotating WD with a period of 33.08 s
($P_{rot} / P_{orb} \approx 0.001$; \citealt{patt79}). The rapid
rotation along with the moderately strong magnetic field of the WD
(1-10 MG) causes almost all of the accreted material ($\sim 99.9$\%;
\citealt{mein15}) to be expelled from the binary in a so called
``magnetic propeller''.  The WD's rapidly spinning magnetosphere acts
as a centrifugal barrier to the accreted diamagnetic plasmoids and
propels them out of the system \citep{wynn95, wynn97}. This accretion
scenario is similar to that of accreting millisecond pulsars.

AE~Aqr also has several other unique observational characteristics.
\citet{deja94} using 14 years of X-ray data found that the WD is
spinning down at a rate of
$\dot{P}_\star \approx 5.64 \times 10^{-14}$ s s$^{-1}$. This rate
equates to a spin-down luminosity
$L_{s-d} = I\Omega\dot{\Omega} \sim 10^{34}$ erg s$^{-1}$. This is
roughly 500 times larger than the accretion luminosity of
$2 \times 10^{31}$ erg s$^{-1}$ \citep{erac96}.

\citet{book87} and \citet{bast88} using the National Radio Astronomy
Observatory (NRAO) Very Large Array found that AE~Aqr is a persistent
radio emitter showing rapid variablity and large flares.  Observations
at 1.5, 4.9, 15, and 22.5 GHz show a long-term averaged spectral index
$\alpha \propto 0.5$ ($S_\nu \propto \nu^\alpha$;
\citealt{bast88}). The flux density at 15 GHz showed variations of a
few mJy on minute timescales during quiescence and peaked at 12 mJy
during flares, which occur about every five hours. They also note that
in those cases where the flares are temporally resolved, the rise time
is shorter than the decay time, which suggests that an external medium
limits the flare's rate of expansion. Ground-based millimeter and
submillimeter, and spaced-based infrared observations show that the
positive spectral index extends to terahertz frequencies
\citep{abad05, dubu07, tork13}. Above 2 THz the spectrum turns over
and has a spectral index of $\alpha \propto -0.7$. This nonthermal
radio emission, spanning more than three orders of magnitude in
frequency, is attributed to a superposition of synchrotron emitting
plasmoids that expand and cool radiatively through synchrotron
radiation \citep{vdla63, vdla66}.

\citet{thor20} recently argued that the CV LAMOST J024048.51+195226.9,
also known as CRTS J024048.5+19227, (hereafter J0240+19) may be a
possible twin of AE~Aqr based on its unusual optical spectrum, which
shows no He II emission and weak He I lines. In addition, the light
curves show large irregular flaring on the timescale of
minutes. However, the time resolution of the photometry was
insufficient to search for pulsations on the order of tens of
seconds. Using publically available Catalina Real-Time Transient
Survey (CRTS) and All-Sky Automated Survey for Supernovae (ASAS-SN)
data, \citet{litt20} identify a shallow dip in the light curves at the
secondary's inferior conjunction and identify this feature with an
eclipse by the secondary star. This means that J0240+19 is a high
inclination system and explains why the He I $\lambda 6678$ \AA\
emission line briefly disappears at orbital phase 0. Additional
photometric observations by \citet{garn21} confirmed the high
inclination of J0240+19, but did not detect the WD's spin period. The
spin period was finally found by \citet{peli21} to be 24.9328(38) s
with an amplitude of 0.2\% in the \textit{g} band, which is below the
detection limits of the previous searches. J0240+19 is now the fastest
known rotating WD. A search of the NRAO Very Large Array Sky Survey
(VLASS) version 1.2 (tile T15t04, 2019 Jun 07; \citealt{lacy20})
showed the presence of a $508 \pm 120\ \mu$Jy ($4\sigma$) radio source
at 3 GHz within 1 arcsec of the J0240+19 position. It was also
detected in a MeerKAT pointed observation on 2020 Aug 12 with a flux
density of $600 \pm 20\ \mu$Jy at 1.284 GHz and spectral index
$\alpha \approx -0.6$ \citep{pret21}. Neither observation reported any
polarization measurements.

In this paper we present three hours of NRAO \textit{Karl G. Jansky}
Very Large Array (VLA) observations of J0240+19. The observations are
split between one hour of spectroscopy in order to compare the
spectrum of J0240+19 to that of AE~Aqr and two hours of photometry in
order to detect the spin period of the WD. If J0240+19 is a twin of
AE~Aqr, then its radio flux density is expected to increase with
frequency and to be $\approx~1.4$ mJy at 22 GHz, assuming a flux
density of $508\ \mu$Jy at 3 GHz and a spectral index of 0.5, the same
as AE~Aqr. At this flux density, it should be possible to detect a
large amplitude pulsation with a period as short as 30 s.

\section{Observations and Data Reduction}

J0240+19 was observed by the VLA for three hours on two seperate
dates. The first observation measured the spectrum of J0240+19 across
five radio bands from 2--26 GHz during a one hour scheduling
block. The exposures in each band range between 600--628 s
duration. The results of this observation determined which frequency
band to use for the second photometric observation of two hours. This
second observation was done using the C band (4--8 GHz) with an
on-source time of about 1.6 hours. During the second observation, the
VLA cycled between the target J0240+19 and the phase calibrator
J0238+1636 every 570 s with about 525 s being on target. Each
observation used the 3-bit samplers for wideband coverage. The radio
sources J0137+3309 (=3C48) and J0238+1636 were used as the flux and
polarization, and the phase and gain calibrators for each observation,
respectively. No polarization leakage calibrators were observed,
because the cross-polarization is $<1$\% and varies slowly over
sevaral months \citep{perl14}. This accuracy is sufficient for our
needs. Note that the flux calibrator 3C48 has been undergoing a flare
since about January 2018, which will affect the absolute flux
scale. The effect is smaller at low frequencies ($\sim 5$\% at S band)
and greater at high frequencies ($\sim 10$\% at Ku band;
\citealt{vsus21}). Table 1 gives a log of the observations.

\begin{deluxetable}{ccccc}
  \tablenum{1}
  \tablecaption{Log of the observations of J0240+19.}
  \tablehead{
    \colhead{Start Date} & \colhead{Start Time} & \colhead{Band} &
    \colhead{Frequency} & \colhead{Exposure} \\
    \colhead{}                 & \colhead{(UTC)}        & \colhead{} &
    \colhead{(GHz)} & \colhead{(h:m:s)}
  }
  \startdata
  2021 Jun 11 & 16:52:36 & K   & 18--26 & 00:05:28 \\
                      & 16:59:54 & Ku & 12--18 & 00:05:22 \\
                      & 17:07:09 & X   &   8--10 & 00:05:12 \\
                      & 17:14:15 & C   &   4--8   & 00:05:10 \\
                      & 17:21:20 & S   &   2--4   & 00:05:00 \\
  2021 Jul 20  & 15:22:50 & C   &   4--8   & 01:35:55 \\
  \enddata
\end{deluxetable}

The data were calibrated using version 6.1.2 of the CASA (Common
Astronomy Applications Software) calibration pipeline
\citep{mcmu07}. The imaging application \textit{tclean} is used to
generate IQUV and RRLL images of each target scan to check for any
source confusion and radio frequency interference. None were
found. The flux densities are measured by fitting a point source to
the UV data using the Julia programming language package
\textit{Visfit} \citep{barr21a}. The package uses a box-constrained
Levenberg-Marquardt algorithm to minimize the model residuals. The
position of the point source is constrained to be within $\pm$2 arcsec
of the known source position. When measuring the left and right
circular polarizations, the flux densities are measured simulaneously
using the same source position. Although the phase centers for the
left and right polarizations may have slightly different positions,
this approach is recommended for faint sources.

\section{Analysis}

\subsection{Spectroscopy}

The spectrum of J0240+19 obtained during the first observation is
shown in Figure 1. This observation occured between orbital phases
0.70--0.78 using the ephemeris of \citet{garn21}.  The average flux
density and standard deviations for each frequency band are shown in
black. It varies from $195 \pm 12$ $\mu$Jy at 3 GHz to $23 \pm 8$
$\mu$Jy at 22 GHZ. The spectral index of these measurements is roughly
consistent with the negative spectral index measured by MeerKAT of
$\alpha \propto -0.6$ \citep{pret21}, where
$S_\nu \propto \nu^\alpha$. Figure 1 also shows the left (LL) and
right (RR) circular polarizations for a narrower bandwidth (512 MHz)
across the entire frequency range. These data show that the circular
polarization varies randomly from one frequency to the next. In
particular, the measurement at 4.231 GHz shows strong ($>50$\%)
circularly polarized emission. Note that the Q and U images show no
significant linear polarization ($\lesssim 1$\%).

\begin{figure}[h]
  \plotone{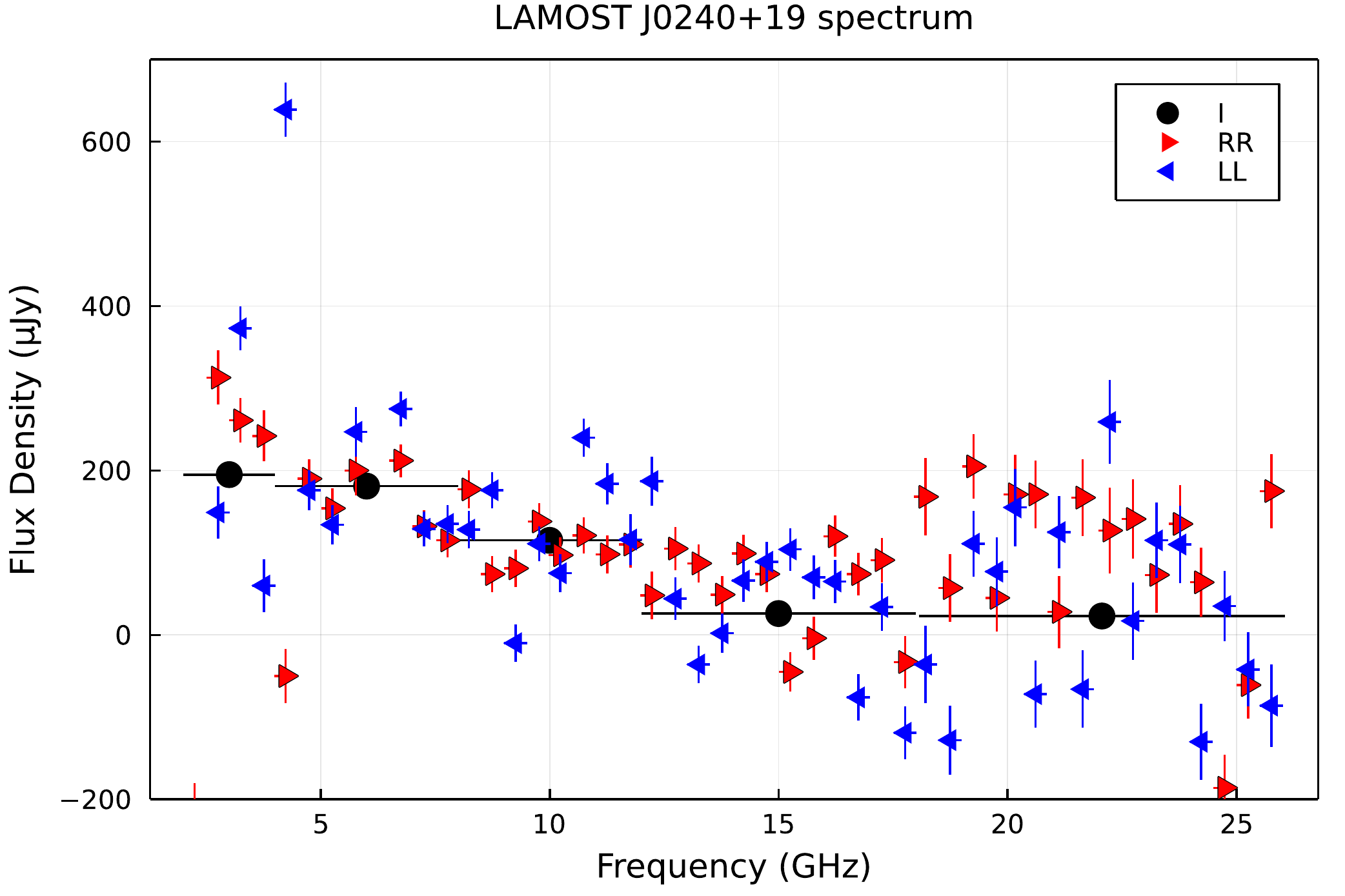}
  \caption{2-26 GHz spectrum of LAMOST J0240+19. The average flux
    densities (black points) are shown for each of the five spectral
    bands (S, C, X, Ku and K). The left (blue) and right (red)
    circular polarizations are also shown for a narrower bandwidth
    (512 MHz). Note the high polarization at 4.231 GHz.}
\end{figure}

\subsection{Photometry}

Because the spectroscopy showed that the S and C bands have similar
flux densities ($\approx~200$ $\mu$Jy), the photometry was done using
the C band, because its wider bandwidth provides greater
sensitivity. Unfortunately, J0240+19 was even weaker during the second
observation with an average flux density of $\approx~100 \mu$Jy. Such
a low flux density precludes any attempt at searching for, or
detecting, a spin period as short as 30 s. We found that the best
compromise between time resolution and signal-to-noise ratio is an
integration time of 265 s.  The observation occurred just after
inferior conjunction between orbital phases 0.08--0.32 using the
ephemeris of \citet{garn21}. The resulting left and right circular
polarization light curves are shown in Figure 2. Like the spectrum
shown in Figure 1, the light curve also shows that the flux density
and circular polarization vary randomly on a timescale of at most a
few minutes. The light curve is too short and variable to show any
evidence of an orbital period.

\begin{figure}[h]
  \plotone{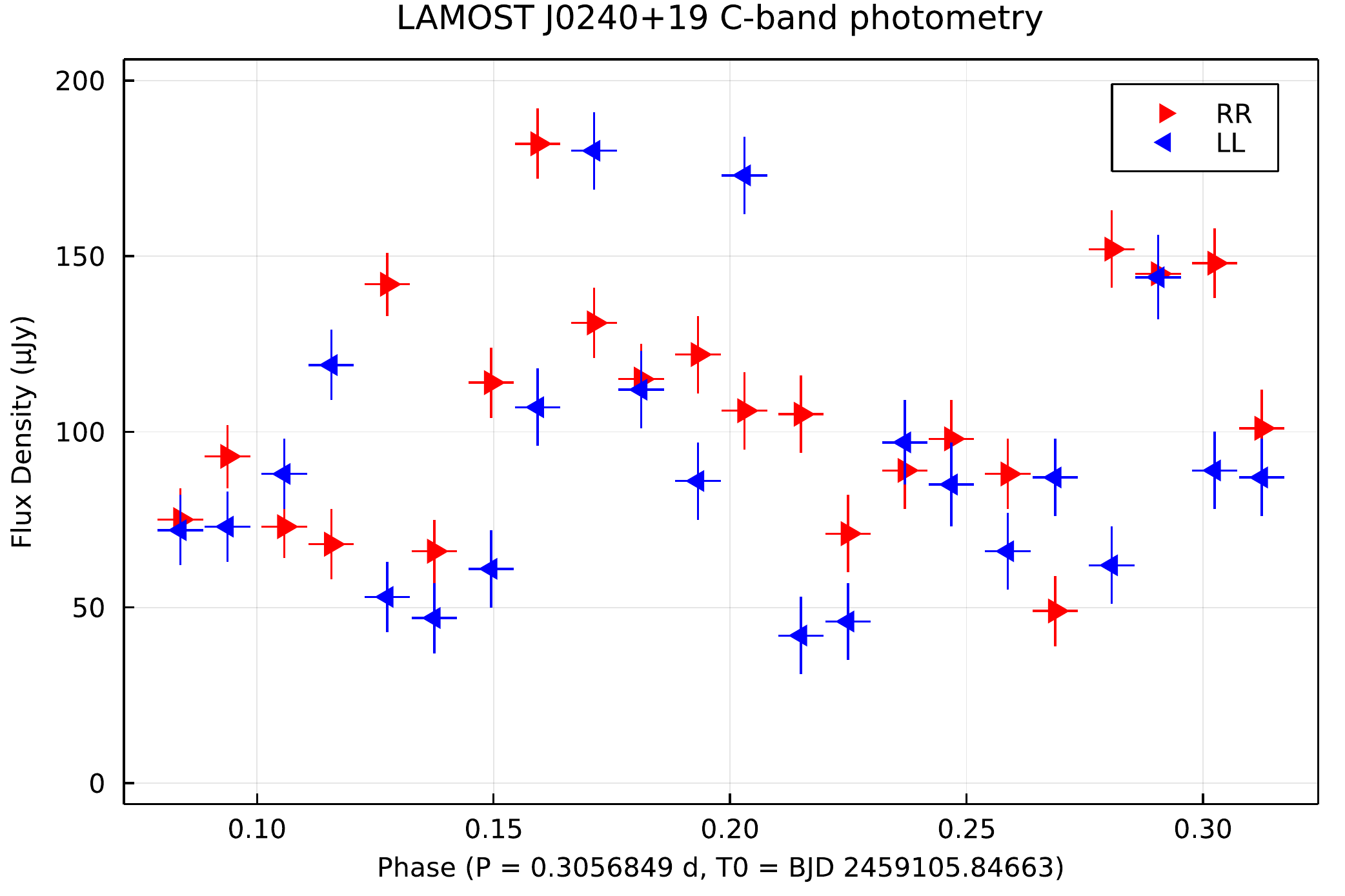}
  \caption{1.6 hours of C band photometry (4--8 GHz) of LAMOST
    J0240+19. The left (blue) and right (red) circular polarization
    are shown. The integration time for each flux measurement is 265
    s.}
\end{figure}

\section{Discussion}

\subsection{Coherent radio emission}

The brightness temperature,
$T_B \approx 10^{14}\ S_{-3} D_{2} \nu_9^{-2} r_{10}^{-2}$ K, where
$S_{-3}$ is the observed flux density in mJy, $D_{2}$ is the distance
in units of 100 pc, $\nu_9$ is the frequency in GHz, and $r_{10}$ is
the radius of the projected source area in $10^{10}$ cm. For J0240+19,
$T_B \approx 10^{13}$ K for $S_{-3} = 0.2$ at 3 GHz, $D = 6.2$
\citep{gaia20}, and $r_{10} = 7$. The value of $r_{10}$ assumes that
the area of the radio emission has a radius twice that of the donor
star. This high $T_B$ implies a coherent emission process.

Following the discussion of \citet{bast87}, there are two coherent
mechanisms that can produce highly circularly polarized emission as
seen in the spectrum and light curve of J0240+19: magnetic plasma
radiation and electron cyclotron maser emission (ECME). Synchrotron
emission as suggested by \citep{pret21} is unlikely. These emissions
can be produced by the ordinary (O), the extraordinary (X), or the low
frequency branch of the extraordinary (Z) modes depending on the
emission mechanism. In the case of plasma emission, the radiation is
produced at either the fundamental or the second harmonic of the
electron plasma frequency, $\omega_{pe} = (4 \pi n_e e^2/ m_e)^{1/2}$,
where $n_e$ is the electron density, $e$ is the electron charge, and
$m_e$ is the electron mass. Emission at the fundamental frequency can
be completely polarized and is ascribed to the O mode. Whereas,
emission at the second harmonic is moderately polarized and is related
to either the O or X modes. For the frequency range of 2--18 GHz,
$\omega_{pe}$ gives a density of the source region of
$\approx 10^9 - 10^{11}$ cm$^{-3}$.

In the case of ECME, the polarized emission is related to the O, X, or
Z modes. Emission at the fundamental gyrofrequency,
$\omega_{Be} = e B / m_e c$, is unlikely to escape the source region,
because of the strong gyroresonant absorption at the second
harmonic. However, emisson at the second harmonic of the O or X modes
may escape the source region and be highly polarized.  The polarized
emission at $\approx 2$ GHz places an upper limit of
$\lesssim 4 \times 10^{10}$ cm$^{-3}$ on the electron density, since
the ECME requires the plasma frequency to be less than the
gyrofrequency. It also implies that the ambient magnetic field of the
source region $B$
($\approx 360\ (\frac{\omega_{Be}}{1\ \small{\textrm{GHz}}})$ G) is in
the range of 360 -- 3200 G.







\subsection{The radio emission compared to AE~Aqr}


The two physical constraints of a low density plasma and kiloGauss
magnetic field restricts the location of the radio emission to a low
density plasma in the magnetosphere of the WD or the lower corona of
the donor star. Emission from the accretion stream or disk is
therefore unlikely because the densities in those regions are too high
($\gtrsim 10^{14}$ cm$^{-3}$ \citealt{barr20}). We first consider
emission from the WD magnetosphere. Assuming a 1 MG dipolar surface
field for the WD, The strength of the magnetic field with radius is:
$B = 10^6 r_9^{-3}$ G, where $r_9$ is the WD radius in $10^9$
cm. Therefore, the radio emission comes from a region between 7--14 WD
radii for magnetic fields of 360--3200 G. These radii are much smaller
than the radius of the donor star and are therefore likely to be
modulated by the orbital period and eclipsed by the donor star.
Although the radio observations are sparse and do not cover the
orbital phase of the eclipse (0.95--0.05), neither the MeerKAT L band
\citep{pret21} nor VLA C band observations show any evidence of an
orbital modulation. The MeerKAT observations occur during orbital
phases 0.25--0.44 and our VLA observation during phases
0.08--0.32. Therefore, emission from the WD magnetosphere is
problematic.

Based on these arguments, the donor star's corona is the most likely
location of the radio emission.  Roche tomography of the donor star of
AE Aqr \citep{dunf05, wats06, smit12} shows that $\sim 20$\% of its
surface is affected by high-latitude starspots similar to that seen in
Doppler images of rapidly rotating isolated stars. This implies that
the donor star in AE Aqr is highly active, which is expected for a
rapidly rotating star ($P_{rot} < 10$ days), because the magnetic
dynamo saturates at about this spin period. We propose that the donor
star in J0240+19 is also magnetically active and is the source of the
radio emission. This means that the surface field of the donor star is
$\sim$ few kG in strength. It also provides an explanation of the
random nature of the circular polarization in time and frequency due
to a superposition of plasma radiation or ECME from several emitting
plasmoids.

\begin{figure}[h]
  \plotone{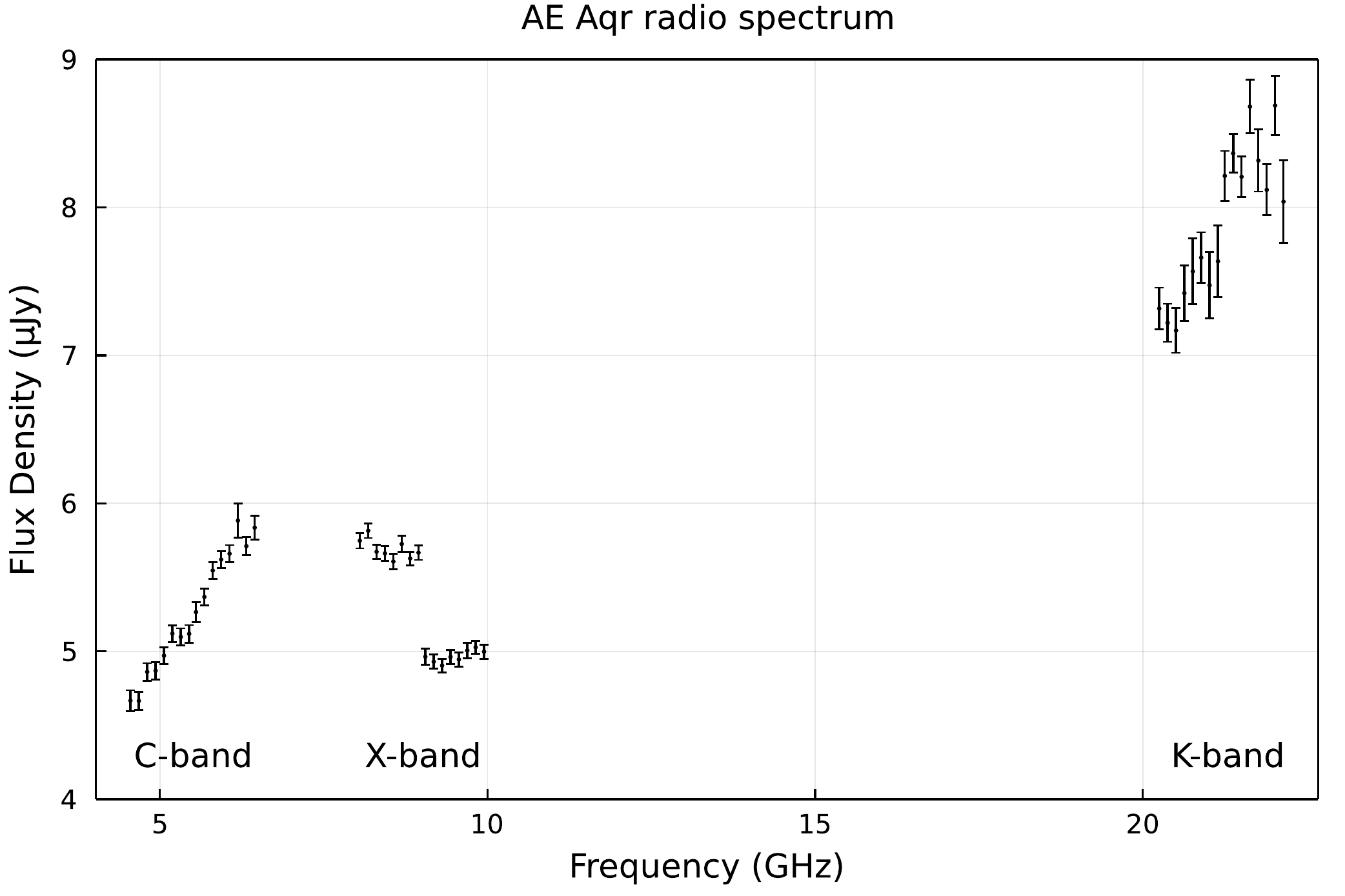}
  \caption{The 4-24 GHz spectrum of AE~Aqr showing a positive spectral
    index betwen the C and K bands.}
\end{figure}

In order to compare the spectra of AE~Aqr and J0240+19, we have
reanalyzed one of the observations of AE~Aqr by \citet{barr17} from
2015. Figure 3 shows a positive spectral index between the C and
K bands (6 GHz and 22 GHz, resp.). Because each band was observed
several minutes apart and the radio emission varies rapidly
($\sim$minutes), the X band (8 GHz) was caught during a period of
negative spectral index. As noted previously, AE~Aqr has a positive
spectral index spanning over three orders of magnitude and only begins
to turnover at $\approx 2$ THz. Also note that radio emission in all
three bands is unpolarized ($P_V \lesssim 0.1$\%). This spectrum is
unlike that of the spectrum of J0240+19 with its negative spectral
index and strong circular polarization. Instead, the spectrum of
J0240+19 is similar to that of the nova-like V603 Aql, which on
average has a negative spectral index between 2--12 GHz and shows
strong, randomly varying, circular polarization \citep{copp15, barr17,
  barr20, barr21b}\footnote{Although, V603 Aql has been suspected of
  being an IP, to date no optical or X-ray spin period has been
  detected (see e.g., \citealt{haef85, udal89, gned90})}.  In
addition, the specific radio luminosities at 3 GHz are of similiar
magnitude. That of J0240+19 is 7.3--18.3 $\times 10^{15}$ ergs
s$^{-1}$ Hz$^{-1}$ and that of V603 Aql is 2.0--3.5 $\times 10^{15}$
ergs s$^{-1}$ Hz$^{-1}$. These luminosities are in the range of 1--50
$\times 10^{15}$ ergs s$^{-1}$ Hz$^{-1}$, which is typical of most
CVs (See Figure 2 of \citealt{pret21}).
 





\section{Conclusions}

The three hours of VLA observations of J0240+19 show a spectrum and
light curve that are dissimilar to those of AE~Aqr. The spectrum has a
negative spectral index between 2--26 GHz and the emission shows
periods of high circular polarization, while those of AE~Aqr show a
positive spectral index and unpolarized emission. The radio emission
from J0240+19 is characteristic of magnetic plasma radiation or
electron cyclotron maser emission, while that of AE~Aqr and AR~Sco is
characteristic of synchrotron emission. Therefore, the radio emission
from J0240+19 is unlike that of AE~Aqr and AR~Sco, although the
characteristics of their optical emission are very similar. The radio
emission from J0240+19 is similar to most radio-emitting magnetic CVs
and is most similar to the nova-like CV V603 Aql, which is suspected
of containing a rapidly spinning WD.

We argue that the source of the radio emission in J0240+19 is the
magnetically active donor star and not a low density region of the
magnetosphere near the WD, because the radio light curves show no
evidence of an orbital modulation. If the radio emission is from the
WD magnetosphere, then it should be eclipsed during inferior
conjunction, because of its proximity to the WD. We therefore
encourage additional radio observations to refute this assertion by
looking for radio eclipses. Although J0240+19 is a weak radio source
($\lesssim 1$ mJy), it is a persistent source in a high-inclination,
eclipsing CV, making it a good laboratory for studying radio emission
from CVs.

\section{Acknowledgements}

We thank Paul Mason for a critical reading of the manuscript and
several useful comments. The National Radio Astronomy Observatory is a
facility of the National Science Foundation operated under cooperative
agreement by Associated Universities, Inc.

\facilities{VLA}

\software{CASA (6.1.2), Julia (1.6.2), Visfit}

\end{document}